\documentclass[preprint,aps,nofootinbib,11pt]{revtex4-1}
\usepackage{geometry}                
\usepackage{amsmath,amssymb,amsfonts,dcolumn,color,graphicx,graphics,latexsym,placeins,epsfig,tikz}
\usetikzlibrary{arrows,shapes}
\usepackage{subfigure,rotating,bm,mathrsfs}
\usepackage[title]{appendix}

\newcommand{\be}{\begin{equation}}
\newcommand{\ee}{\end{equation}}
\newcommand{\ba}{\begin{eqnarray}}
\newcommand{\ea}{\end{eqnarray}}

\usepackage[pagebackref=false, colorlinks=true]{hyperref}
\definecolor{redish}{rgb}{0.7,0.2,0.0}  
\definecolor{bluish}{rgb}{0.2,0.5,0.8}

\hypersetup{linkcolor=redish,          
                  citecolor=blue,        
                  filecolor=magenta,      
                  urlcolor=blue}          

\begin{document}

\author{Rajibul Shaikh}
\email{rshaikh@iitk.ac.in}
\author{Kunal Pal}
\email{kunalpal@iitk.ac.in}
\author{Kuntal Pal}
\email{kuntal@iitk.ac.in}
\author{Tapobrata Sarkar}
\email{tapo@iitk.ac.in}
\affiliation{Department of Physics, \\ Indian Institute of Technology, \\ Kanpur 208016, India}

\title{\Large Constraining alternatives to the Kerr black hole}

\begin{abstract}

The recent observation of the shadow of the supermassive compact object M87$^*$ by the Event Horizon Telescope (EHT) collaboration has opened up a new window to probe the strong gravity regime. In this paper, we study shadows cast by two viable alternatives to the Kerr black hole, and compare them with the shadow of M87$^{*}$. The first alternative is a horizonless compact object (HCO) having radius $r_0$ and exterior Kerr geometry. The second one is a rotating generalisation of the recently obtained one parameter ($r_0$) static metric by Simpson and Visser. This latter metric, constructed by using the Newman-Janis algorithm, is a special case of a parametrised rotating non-Kerr geometry obtained by Johannsen. Here, we constrain the parameter $r_0$ of these alternatives using the results from M87$^{*}$ observation. We find that, for the mass, inclination angle and the angular diameter of the shadow of M87$^{*}$ reported by the EHT collaboration, the maximum value of the parameter $r_0$ must be in the range $2.54r_{+}\leq r_{0,max}\leq 3.51r_{+}$ for the dimensionless spin range $0.5\leq a_{*}\leq 0.94$, with $r_{+}$ being the outer horizon radius of the Kerr black hole at the corresponding spin value. We conclude that these black hole alternatives having $r_0$ below this maximum range (i.e. $r_0\leq r_{0,max}$) is consistent with the size and deviation from circularity of the observed shadow of M87$^{*}$.

\end{abstract}

\maketitle
\section{Introduction}

Understanding the nature of strong gravity is perhaps the key to the formulation of a full theory of quantum gravity that
remains elusive more than a hundred years after Einstein formulated the theory of general relativity (GR). 
Ubiquitous in such studies are black holes and the associated null surfaces, i.e., event horizons. It is now believed that
the centers of most galaxies contain supermassive black holes with masses of the order of $10^6~-~10^{10}M_{\odot}$.
An important observational signature of the event horizon of a black hole is its shadow -- a dark region surrounding 
the central singularity, caused by gravitational lensing, and the capture of photons at the horizon due to strong gravity.  
Phenomenal advances in the observational study of black hole event horizons have come about recently, after
the first results on the radio source M87$^{*}$ were published by the Event Horizon Telescope (EHT) collaboration
\cite{EHT1},\cite{EHT2},\cite{EHT3}. 
These results have opened up the exciting possibility of understanding strong gravity near black hole event horizons,
which, even a few years back, seemed like a purely theoretical aspect. 

While the phenomenology of black holes continues to be at the focus of attention post the EHT results, 
horizonless compact objects (HCOs) are fast gaining popularity in the literature. Part of the reason for 
studying such objects is that these can be black hole alternatives, whose observational aspects can 
be similar to those of black holes. Indeed, since the existence of a singularity 
usually indicates a pathology in a theory, the existence of black hole singularities in a purely classical version of GR
is still debated, with many advocating that it is removed by quantum effects, a result that is well known in 
semi classical versions of GR. 

Further, as is well known by now, HCOs, which 
do not have an event horizon, might mimick many properties of black holes.  
Hence, in this EHT era, it becomes all the more important to study black hole mimickers, as obseravational 
signatures from these can be contrasted and compared with the EHT results on M87$^{*}$. 
For example, one such black hole mimicker is the wormhole solution of GR, which contains a throat that connects 
two different universes or two distant regions of one universe. Although the presence of the throat indicates the breakdown 
of the weak energy condition, it is well known that in dynamical scenarios, or for wormholes in modified gravity, such 
energy condition violations can be avoided. In \cite{DamourSolodukhin}, Damour and Solodukhin 
pointed out the close similarities of several theoretical features of black holes with those obtained from wormhole
solutions. Apart from wormholes, various other possible HCO geometries have been studied in the literature,
see, e.g., \citep{CardosoReview} and the references therein.

An important aspect of the EHT data is that it can be used to constrain the parameter space of geometries which deviate from Schwarzschild or Kerr black hole (see, e.g., \cite{Psaltis}, \cite{Soumitra}, \cite{Bambi}, \cite{Rahul}). In this paper, we use this approach to put bounds on two Kerr black hole alternatives. The first alternative is a horizonless compact object (HCO) having radius $r_0$ and exterior Kerr geometry. For the second one, we construct a metric by generalising a recently proposed static, spherically symmetric solution of
Einstein's equation, given by Simpson and Visser (SV) \cite{SV1}. The SV metric is attractive 
in that it is a minimal one parameter extension of the Schwarzschild metric, and can describe a black hole
or a wormhole for different choices of the parameter. Here, a rotating version of the SV metric is first derived using the 
Newman-Janis algorithm, 
which can correspond to a rotating black hole or a rotating wormhole. Using these two alternatives, we 
study the shadow and compare it with the data for the shadow of  M87$^{*}$. 
In the process, we are able to put a bound on the parameter appearing in the solutions. 

This paper is organised as follows. In Sec. \ref{sec:metric}, we discuss the spacetime geometry of the black hole alternatives. In Sec. \ref{sec:construction}, we discuss the detailed method for constructing the shadows of the alternatives. We constrain the parameter of these black hole alternatives using the M87$^*$ results in Sec. \ref{sec:results}. We conclude in Sec. \ref{sec:conclusion}. 
This paper also has an appendix where the relevant details of the Newman-Janis algorithm is discussed briefly. \\

\noindent
{\bf Note added :} While this paper was being readied for submission, we became aware of the work of \cite{Liberati},
where the authors derive the rotating SV solution and study in details the resulting phase diagram. 

\section{The Kerr black hole and its alternatives}
\label{sec:metric}
\subsection{Kerr black hole}
The Kerr metric in Boyer-Lindquist coordinates can be written as
\begin{equation} 
ds^2=-\left(1-\frac{2Mr}{\Sigma}\right)dt^2-\frac{4Mar \sin^2\theta}{\Sigma}dt d\phi +\frac{\Sigma}{\Delta}dr^2 +\Sigma d\theta^2+\left( r^2+a^2+\frac{2Ma^2r \sin^2\theta}
{\Sigma}\right) \sin^2\theta d\phi^2,
\label{kerr}
\end{equation}
where $M$ is the mass of the black hole, $a$ is the specific angular momentum defined as $a=J/M$ and
\begin{equation}
 \Sigma=r^2+a^2 \cos^2\theta, \hspace{0.5cm}   \Delta=r^2-2Mr+a^2 .
\end{equation}
For convenience, we define a dimensionless Kerr parameter called spin as $a_{*}=a/M=J/M^{2}$. The horizon radii of the black hole are the roots of $\Delta=0$ and are given by
\begin{equation}
r_{\pm} =
M(1\pm\sqrt{1-a_{*}^2}).
\end{equation}

\subsection{Horizonless compact objects with exterior Kerr metric}
We consider a HCO with the exterior being given by the Kerr metric, and its surface outside the outer horizon of the would-be Kerr black hole, i.e., at $r=r_0>r_+$, $r_+$ being the outer horizon radius of the Kerr black hole.

\subsection{Rotating Simpson-Visser metric: a special case of the Johannsen metric}

Recently, Simpson and Visser constructed a metric where the central singularity is replaced by a minimal surface of radius $r_0$ and can acts as a black hole mimicker. The spacetime geometry is given by \citep{SV1}
\begin{equation}
ds^2=-\left(1-\frac{2M}{\sqrt{r^2+r_0^2}}\right)dt^2+\frac{dr^2}{1-\frac{2M}{\sqrt{r^2+r_0^2}}}+(r^2+r_0^2)(d\theta^2+\sin^2\theta d\phi^2).
\end{equation}
The above geometry represents a regular black hole when $r_0<2M$ and a wormhole when $r_0>2M$ with $r_0$ being the wormhole throat radius. In the coordinate system used above, the throat corresponds to $r=0$. We now apply the Newman-Janis algorithm to obtain a rotating version of the above metric. After doing this the metric becomes (see Appendix \ref{sec:appendix})
\begin{eqnarray} 
ds^2 &=&-\left(1-\frac{2M\sqrt{r^2+r_0^2}}{\Sigma}\right)dt^2-\frac{4Ma\sqrt{r^2+r_0^2} \sin^2\theta}{\Sigma}dt d\phi +\frac{\Sigma}{\Delta}dr^2 \nonumber \\
& & +\Sigma d\theta^2 + \left( r^2+r_0^2+a^2+\frac{2Ma^2\sqrt{r^2+r_0^2} \sin^2\theta}
{\Sigma}\right) \sin^2\theta d\phi^2,
\end{eqnarray}
\begin{equation}
 \Sigma=r^2+r_0^2+a^2 \cos^2\theta, \quad   \Delta=r^2+r_0^2-2Mr+a^2.
\end{equation}
Note that it reduces to Kerr geometry when $r_0=0$. While this work was in preparation, in \citep{Liberati} which appeared recently on arXiv, the authors have also obtained the above metric through the same method as ours.

The above metric can further be simplified using the coordinate transformation $\bar{r}=\sqrt{r^2+r_0^2}$. Performing the transformation and dropping the bar, we obtain
\begin{eqnarray} 
ds^2 &=&-\left(1-\frac{2Mr}{\Sigma}\right)dt^2-\frac{4Mar \sin^2\theta}{\Sigma}dt d\phi +\frac{\Sigma}{\Delta\hat{\Delta}}dr^2 \nonumber \\
& & +\Sigma d\theta^2 + \left( r^2+a^2+\frac{2Ma^2r \sin^2\theta}
{\Sigma}\right) \sin^2\theta d\phi^2,
\label{kerrWH}
\end{eqnarray}
\begin{equation}
 \Sigma=r^2+a^2 \cos^2\theta, \quad   \Delta=r^2-2Mr+a^2, \quad \hat{\Delta}=1-\frac{r_0^2}{r^2},
\end{equation}
which is very similar to the Kerr geometry except the term $\hat{\Delta}$. For $r_0=0$, $\hat{\Delta}=1$ and we have the Kerr black hole with the horizon radii given by $r_\pm$. Note that, the above metric has also horizons at $r_{\pm}$. However, depending on the values of $r_0$ and the spin $a$, the horizons may or may not be relevant. For $0\leq a/M\leq 1$, the above geometry represents black hole when $r_0<r_{+}$ and wormhole when $r_0\geq r_{+}$. However, for $a/M>1$, it always represents wormhole as $r_{\pm}$ does not exist in this case. In case of wormhole, the throat is given by $\hat{\Delta}=0$ and is at $r=r_0$.

It can be shown that the above rotating Simpson-Visser (SV) belongs to a special case of the parametrized non-Kerr metric constructed by Johannsen. The non-Kerr metric by Johannsen \cite{Johannsen} is given by
\begin{eqnarray}
g_{tt} &=& -\frac{\tilde{\Sigma}[\Delta-a^2A_2(r)^2\sin^2\theta]}{[(r^2+a^2)A_1(r)-a^2A_2(r)\sin^2\theta]^2}~,~~
g_{t\phi} = -\frac{a[(r^2+a^2)A_1(r)A_2(r)-\Delta]\tilde{\Sigma}\sin^2\theta}{[(r^2+a^2)A_1(r)-a^2A_2(r)\sin^2\theta]^2}~, \nonumber \\
g_{rr} &=& \frac{\tilde{\Sigma}}{\Delta A_5(r)}~,~~ g_{\theta \theta} =\tilde{\Sigma}~,~~ 
g_{\phi \phi} = \frac{\tilde{\Sigma} \sin^2 \theta \left[(r^2 + a^2)^2 A_1(r)^2 - a^2 \Delta \sin^2 \theta \right]}{[(r^2+a^2)A_1(r)-a^2A_2(r)\sin^2\theta]^2},
\label{eq:metric}
\end{eqnarray}
where
\begin{eqnarray}
A_1(r) &=& 1 + \sum_{n=3}^\infty \alpha_{1n} \left( \frac{M}{r} \right)^n~,~~
A_2(r) = 1 + \sum_{n=2}^\infty \alpha_{2n} \left( \frac{M}{r} \right)^n~,~~
A_5(r) = 1 + \sum_{n=2}^\infty \alpha_{5n} \left( \frac{M}{r} \right)^n, \nonumber\\
\tilde{\Sigma} &=& \Sigma + f(r)~,~~f(r) = \sum_{n=3}^\infty\epsilon_n \frac{M^n}{r^{n-2}}.
\label{eq:f}
\end{eqnarray}
The Kerr metric is recovered when $f(r)=0$, $A_1(r)=A_2(r)=A_5(r)=1$, i.e., $\alpha_{1n}=\alpha_{2n}=\alpha_{5n}=\epsilon_{n}=0$. The rotating SV metric in Eq. (\ref{kerrWH}) can be obtained as a special case of the above Johannsen metric for $f(r)=0$, $A_1(r)=A_2(r)=1$ and $A_5(r)=\hat{\Delta}(r)$, i.e., for
\begin{equation}
\alpha_{1n}=\alpha_{2n}=\epsilon_{n}=0, \quad \alpha_{5n}=0\;(n\neq 2), \quad \alpha_{52}=-\frac{r_0^2}{M^2}.
\end{equation}

\section{Constructing shadows of the black hole alternatives}
\label{sec:construction}
\subsection{Kerr black hole}
First, we consider a Kerr black hole. The separated geodesic equations which we need for the purpose of shadow in this case are given by
\begin{equation}
\Sigma\frac{dr}{d\lambda}=\pm \sqrt{R(r)}~,~~
\Sigma\frac{d\theta}{d\lambda}=\pm \sqrt{\Theta(\theta)},
\label{eq:theta_eqn}
\end{equation}
where
\begin{equation}
R(r)=\left[(r^2+a^2)E-aL\right]^2-\Delta\left[\mathcal{K}+\left(L-aE\right)^2\right]~,~~
\Theta(\theta)=\mathcal{K}+a^2E^2\cos^2\theta-L^2\cot^2\theta,
\end{equation}
$E$ is the energy, $L$ is the angular momentum, and $\mathcal{K}$ is the Carter's constant.

The unstable circular photon orbits, which form the boundary of a shadow, are given by $\dot{r}=0$, $\ddot{r}=0$ and $\dddot{r}>0$, i.e., by
\begin{equation}
R(r_{ph})=0,\quad R'(r_{ph})=0, \quad R''(r_{ph})>0,
\label{eq:R_condition1}
\end{equation}
where $r_{ph}$ is the radius of an unstable photon orbit. For the Kerr spacetime, after using the first two conditions, we obtain the following critical impact parameters
\begin{equation}
\xi_{ph}=\frac{4Mr_{ph}^2-(r_{ph}+M)(r_{ph}^2+a^2)}{a(r_{ph}-M)}~,~~
\eta_{ph}=\frac{4Ma^2r_{ph}^3-r_{ph}^2\left[r_{ph}(r_{ph}-3M)\right]^2}{a^2(r_{ph}-M)^2},
\label{eq:eta}
\end{equation}
where $\xi=L/E$ and $\eta=\mathcal{K}/E^2$. However, the apparent shape of a shadow is measured using the celestial 
coordinates defined by \citep{celestial}
\begin{equation}
\alpha=\lim_{r_o\to\infty}\left(-r_o^2\sin\theta_o\frac{d\phi}{dr}\Big\vert_{(r_o,\theta_o)}\right)~,~~
\beta=\lim_{r_o\to\infty}\left(r_o^2\frac{d\theta}{dr}\Big\vert_{(r_o,\theta_o)}\right),
\end{equation}
where $(r_o,\theta_o)$ are the position coordinates of the observer. Using the geodesic equations, we obtain
\begin{equation}
\alpha=-\frac{\xi}{\sin\theta_o},
\label{eq:alpha}
\end{equation}
\begin{equation}
\beta=\pm \sqrt{\eta+a^2\cos^2\theta_o-\xi^2\cot^2\theta_o},
\label{eq:beta}
\end{equation}
The contour of the shadow which is formed by the unstable photon orbits is given by the parametric plot of $\alpha_{ph}$ and $\beta_{ph}$ where
\begin{equation}
\alpha_{ph}=-\frac{\xi_{ph}}{\sin\theta_o},
\label{eq:alpha_ph}
\end{equation}
\begin{equation}
\beta_{ph}=\pm \sqrt{\eta_{ph}+a^2\cos^2\theta_o-\xi_{ph}^2\cot^2\theta_o},
\label{eq:beta_ph}
\end{equation}

\begin{figure}[h]
\centering
\includegraphics[scale=0.75]{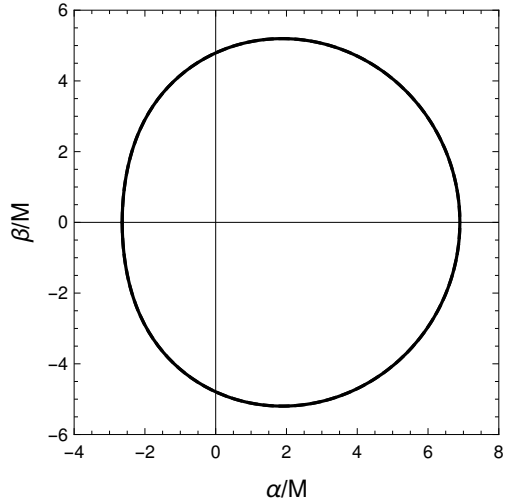}
\caption{Shadow of the Kerr black hole with $a_*=a/M=0.94$ and $\theta_o=\pi/2$.}
\label{fig:shadowKerr}
\end{figure}

For a rotating Kerr black hole, unstable photon orbits, which form the shadow contour in the $\alpha-\beta$ plane, consist of both prograde (photons having motion in the direction of the spin) and retrograde (photons having motion in a direction opposite to the spin) orbits. The prograde orbits have lesser radii than those of the retrograde ones. Figure \ref{fig:shadowKerr} shows a typical shadow of a Kerr black hole. The spin axis divide the shadow into two parts in the $\alpha-\beta$ plane. The contour of the shadow on the left side (i.e., $\alpha<0$ side) of the spin axis is due to the unstable photon orbits which are prograde ($\xi>0$), while the one at the other side (i.e., $\alpha>0$ side) of the spin axis are due to unstable photon orbits which are retrograde ($\xi<0$). Among the unstable photon orbits which form the shadow, the one which has the minimum radius $r_{ph,min}$ is of prograde type and corresponds to the point $(\alpha_{ph,min},0)$ of the shadow contour. On the other hand, the unstable photon orbit which has the maximum radius $r_{ph,max}$ is of retrograde type and corresponds to the point $(\alpha_{ph,max},0)$ of the shadow contour. Therefore, the shadow is formed by the unstable photon orbits whose radii are in the range $r_{ph,min}\leq r_{ph}\leq r_{ph,max}$. The unstable photon orbit radii $r_{ph,max}$ and $r_{ph,min}$ are obtained by $\beta_{ph}=0$, i.e., by
\begin{equation}
\eta_{ph}+a^2\cos^2\theta_o-\xi_{ph}^2\cot^2\theta_o=0.
\end{equation}
$r_{ph,max}$ and $r_{ph,min}$ are respectively given by the largest and the second largest roots of the above equation. Note that besides the mass $M$ and the spin $a$ of the black hole, the radii depend on the observation (inclination) angle $\theta_o$.

\subsection{Horizonless compact object}
Let us now consider an horizonless compact object (HCO) with its surface at $r=r_0$ and with the exterior given by the Kerr solution. If $r_0<r_{ph,min}$, then the compact object is completely hidden inside all the unstable photon orbits of the exterior Kerr metric which take part in the shadow formation. In such a case, the shadow contour of the HCO contour is the same as that of the Kerr black hole. The HCO in this case therefore perfectly mimic a Kerr black hole as far as the shadow silhouette is concerned. However, if $r_0>r_{ph,min}$, then the unstable photon orbits having radii in the range $r_{ph,min}\leq r_{ph}<r_0$ become irrelevant as they lie inside the surface of the compact object. In such a case, the part of the shadow contour, which is lost due to the unstable orbits lying inside the surface, is given by the photons which have turning points at the surface $r=r_0$, i.e., by $R(r_0)=0$. This gives
\begin{equation}
\left[(r_0^2+a^2)-a\xi_0\right]^2-\Delta(r_0)\left[\eta_0+\left(\xi_0-a\right)^2\right]=0,
\label{eq:xi-eta-r0}
\end{equation}
where $\xi_0$ and $\eta_0$ denote the impact parameters of photons having turning points at the surface $r=r_0$. After using Eqs. (\ref{eq:alpha}) and (\ref{eq:beta}), we obtain
\begin{equation}
\left[(r_0^2+a^2)+a\sin\theta_o\alpha_0\right]^2-\Delta(r_0)\left[\beta_0^2+\left(\alpha_0+a\sin\theta_o\right)^2\right]=0,
\label{eq:alpha-beta-0}
\end{equation}
where $(\alpha_0,\beta_0)$ denotes the celestial coordinates of photons having turning points at the surface $r=r_0$. Therefore, for $r_{ph,min}<r_0<r_{ph,max}$, the complete shadow of a HCO is given by the Union of the $(\alpha_{ph},\beta_{ph})$ curve which is due to the unstable photon orbits and the above $(\alpha_0,\beta_0)$ curve. We now describe the details procedure to obtain the shadow. Note that, for $r_{ph,min}<r_0<r_{ph,max}$, $(\alpha_0,\beta_0)$ curve must intersect the $(\alpha_{ph},\beta_{ph})$ curve at the points $(\alpha_{0,max},\pm \beta_{0,max})$ which corresponds to $r_{ph}=r_0$, where $\alpha_{0,max}=\alpha_{ph}\big\vert_{r_{ph}=r_0}$ and $\beta_{0,max}=\beta_{ph}\big\vert_{r_{ph}=r_0}$. Therefore, ($\alpha_0,\beta_0$) in Eq. (\ref{eq:alpha-beta-0}) have the ranges $\alpha_{0,min}\leq \alpha_0\leq \alpha_{0,max}$ and $-\beta_{0,max}\leq \beta_0\leq \beta_{0,max}$. Here, $\alpha_{0,min}$ is obtained by putting $\beta_0=0$ in Eq. (\ref{eq:alpha-beta-0}). This gives
\begin{equation}
\alpha_{0,min}=\frac{r_0^2+a^2\mp a\sin\theta_o\sqrt{\Delta(r_0)}}{\pm\sqrt{\Delta(r_0)}-a\sin\theta_o},
\end{equation}
where we take the root which is negative and close to $\alpha_{0,max}$. Therefore, when $r_{ph,min}<r_0<r_{ph,max}$, the complete shadow contour is given by the union of the $(\alpha_{ph},\beta_{ph})$ curve (due to the unstable photon orbits) plotted for $r_0\leq r_{ph}\leq r_{ph,max}$ and the $(\alpha_0,\beta_0)$ curve (due to photons having turning points at the $r=r_0$ surface) plotted between the points $(\alpha_{0,min},0)$ and $(\alpha_{0,max},\beta_{0,max})$. However, when $r_0>r_{ph,max}$, then the all the unstable photon orbits become irrelevant as all of them lie inside the surface and the shadow in this case is completely given by the $(\alpha_0,\beta_0)$ curve. Note that, when plotting the curve $(\alpha_0,\beta_0)$ [Eq. (\ref{eq:alpha-beta-0})], we simplify it to obtain
\begin{equation}
\beta_0=\pm\frac{\sqrt{\left[(r_0^2+a^2)+a\sin\theta_o\alpha_0\right]^2-\Delta(r_0)\left(\alpha_0+a\sin\theta_o\right)^2}}{\sqrt{\Delta(r_0)}},
\end{equation}
and vary $\alpha_0$ from $\alpha_{0,min}$ to $\alpha_{0,max}$.

\subsection{Rotating Simpson-Visser metric}

Let us now consider the rotating SV metric. The separated geodesic equations are the same except that the radial equation in this case is modified to
\begin{equation}
\Sigma\frac{dr}{d\lambda}=\pm\sqrt{\hat{\Delta}(r)} \sqrt{R(r)}.
\end{equation}
Note that $R(r)$ in this case is the same as that for the Kerr black hole. The construction of shadow contour from the unstable photon orbits is bit involved in this case. Therefore, we consider the black hole and wormhole case separately below. However, to this end, we make use of the unstable photon orbits obtained just from $R(r)$ and its derivative and elaborate what effect the additional factor $\hat{\Delta}(r)$ have in this SV metric case. 

\noindent {\bf I. Black hole case ($r_0<r_{+}$ with $0\leq a/M\leq 1$):} In this case, $r_{ph}>r_0$ always as the unstable photon orbits which take part in shadow formation lie outside the outer event horizon $r_{+}$ and $\hat{\Delta}(r_{ph})\neq 0$ at the location of a unstable photon orbits. Therefore, the conditions $\dot{r}=0$, $\ddot{r}=0$ and $\dddot{r}>0$ for the unstable photon orbits in this case turns out to be $R(r_{ph})=0$, $R'(r_{ph})=0$ and $R''(r_{ph})>0$, which is the same as that for the Kerr black hole. Hence, the unstable circular orbits as well as the shadow contour in this case are the same as those of the Kerr black hole. As a result, this black hole case acts as a perfect Kerr black hole mimicker as far as the shadow silhouette is concerned.

\noindent {\bf II. Wormhole case (either $r_0\geq r_{+}$ with $0\leq a/M\leq 1$ or $a/M>1$):} Depending on the relative values of $r_0$ and the minimum radius $r_{ph,min}$ of the unstable photon orbits obtained from $R(r)$ and its derivative, this case can be subdivided into two subcases.

{\bf IIa:} If $r_0<r_{ph,min}$, then this subcase is similar to the black hole case discussed above. This subcase therefore mimic the Kerr black hole.

{\bf IIb:} If $r_0>r_{ph,min}$, then, for the unstable photon orbits which lies outside the throat, i.e., for $r_{ph}>r_0$, the conditions $\dot{r}=0$, $\ddot{r}=0$ and $\dddot{r}>0$ for them turns out to be $R(r_{ph})=0$, $R'(r_{ph})=0$ and $R''(r_{ph})>0$, which is the same as that for the Kerr black hole. However, it is known that a wormhole throat can acts as a natural position of unstable circular orbits (see \cite{rajibul_2018b} for details). Therefore, for unstable photon orbits which lie at the throat, i.e., for $r_{ph}=r_0$, $\hat{\Delta}(r_{ph})=0$ and $\dot{r}(r_{ph})=0$. Therefore, for such orbits, the conditions for them now become $R(r_0)=0$ and $R'(r_{0})>0$, which are different from those for the unstable orbits lying outside the throat. Note that the condition $R(r_0)=0$ leads to equation (\ref{eq:xi-eta-r0}). Therefore, the shadow contour and the procedure to obtain it for both the horizonless compact object (HCO) with radius $r_0$ and the wormhole in this case will be the same. The only difference in these two cases is that the part of the shadow which gets modified by the $r=r_0$ surface and is given by $R(r_0)=0$ corresponds to photons having turning point at $r=r_0$ for the horizonless object with radius $r_0$, but, for the wormhole, to unstable photon orbits lying at the throat $r_0$.

\begin{figure}[h]
\centering
\subfigure[]{\includegraphics[scale=0.61]{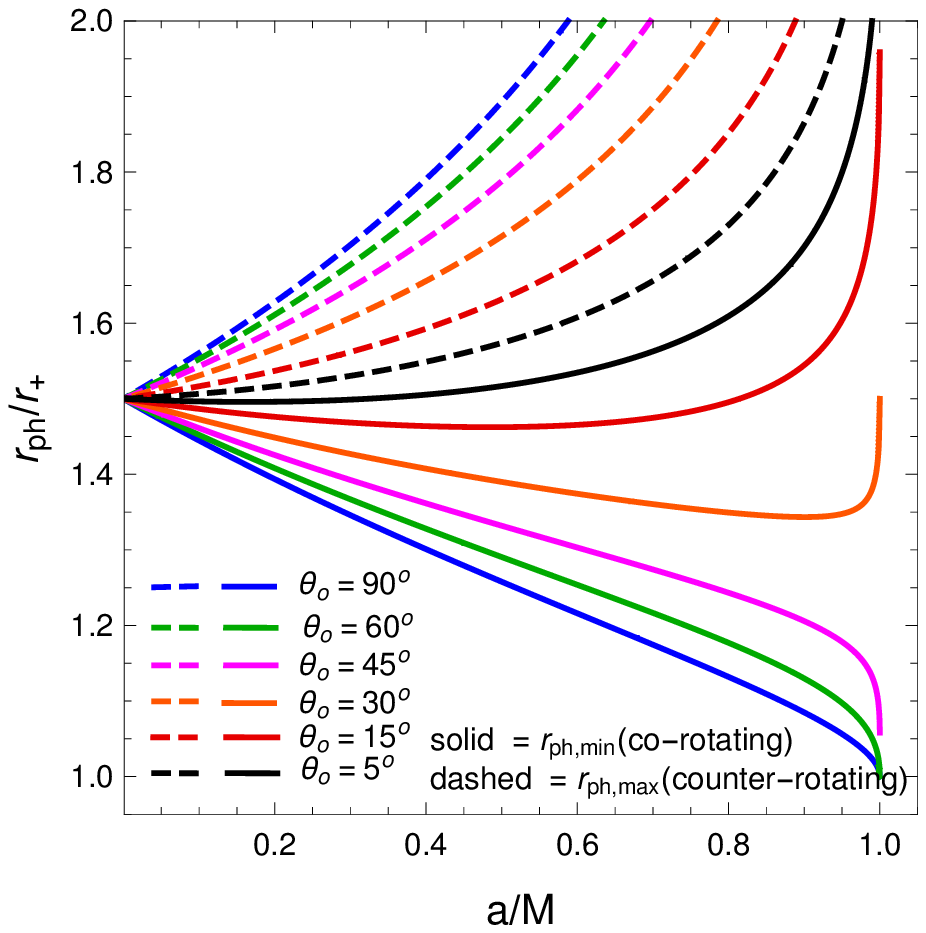}}
\subfigure[]{\includegraphics[scale=0.65]{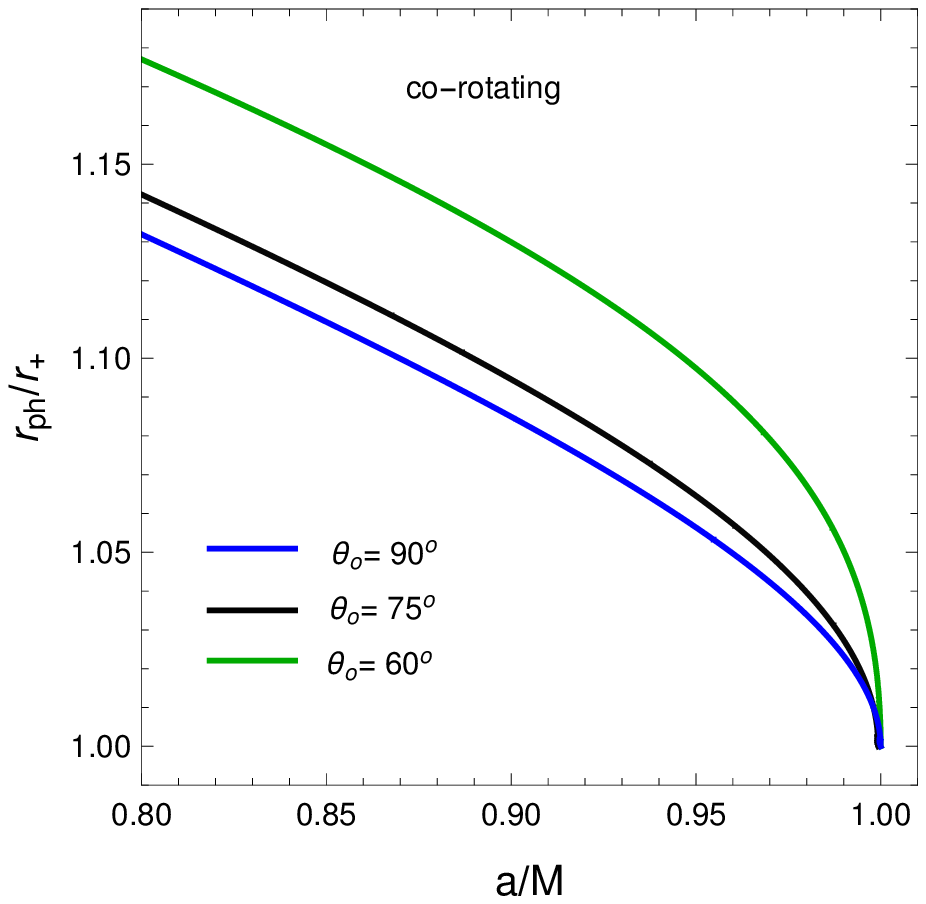}}
\subfigure[]{\includegraphics[scale=0.63]{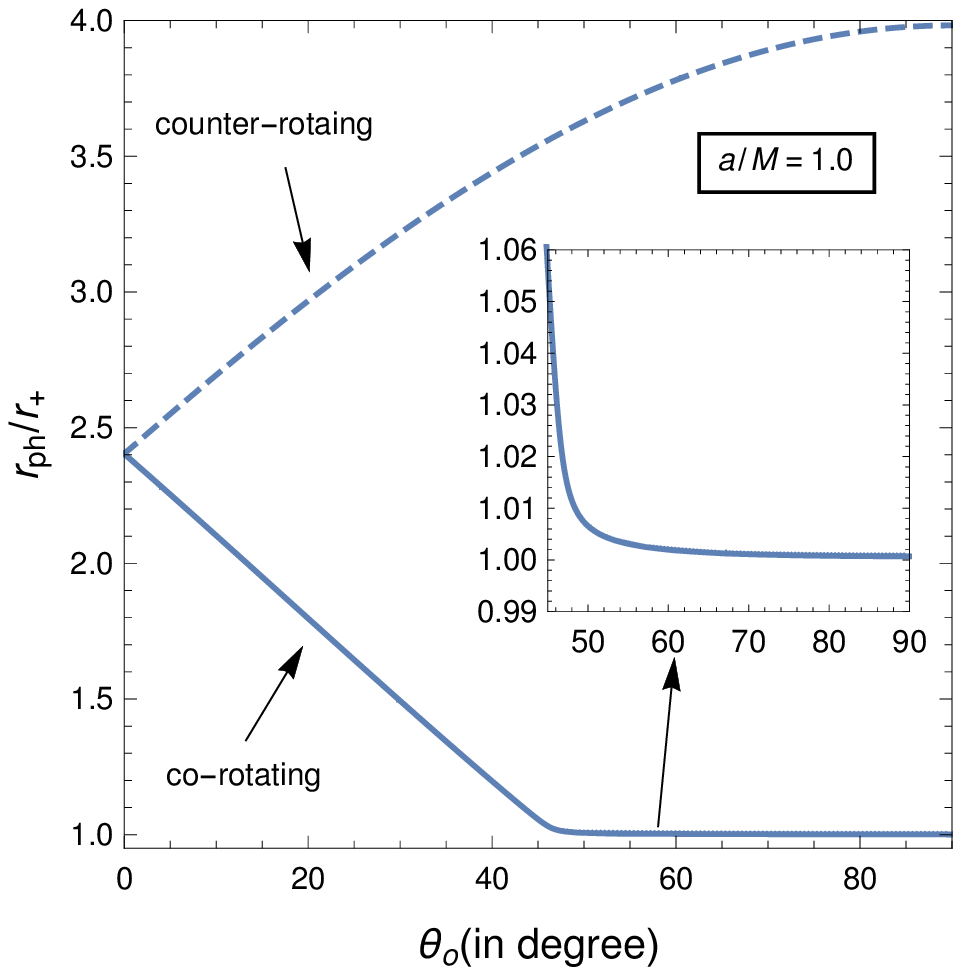}}
\caption{Minimum (solid) and maximum (dashed) radius of the unstable photon orbits which take part in the shadow formation in the Kerr black hole background for a given spin and observer inclination angle. The minimum and the maximum radius correspond to co-rotating and counter-rotating orbits, respectively.}
\label{fig:rph}
\end{figure}

\begin{figure}[h]
\centering
\subfigure[$~r_0/r_{+}=1.05$]{\includegraphics[scale=0.65]{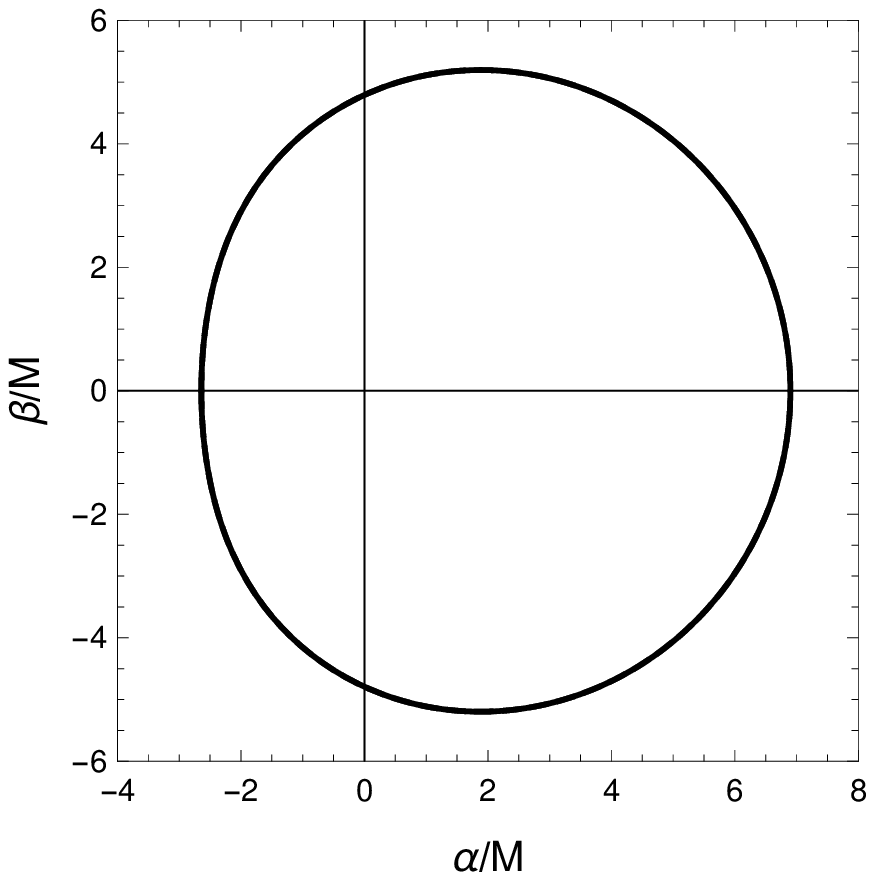}}
\subfigure[$~r_0/r_{+}=1.25$]{\includegraphics[scale=0.65]{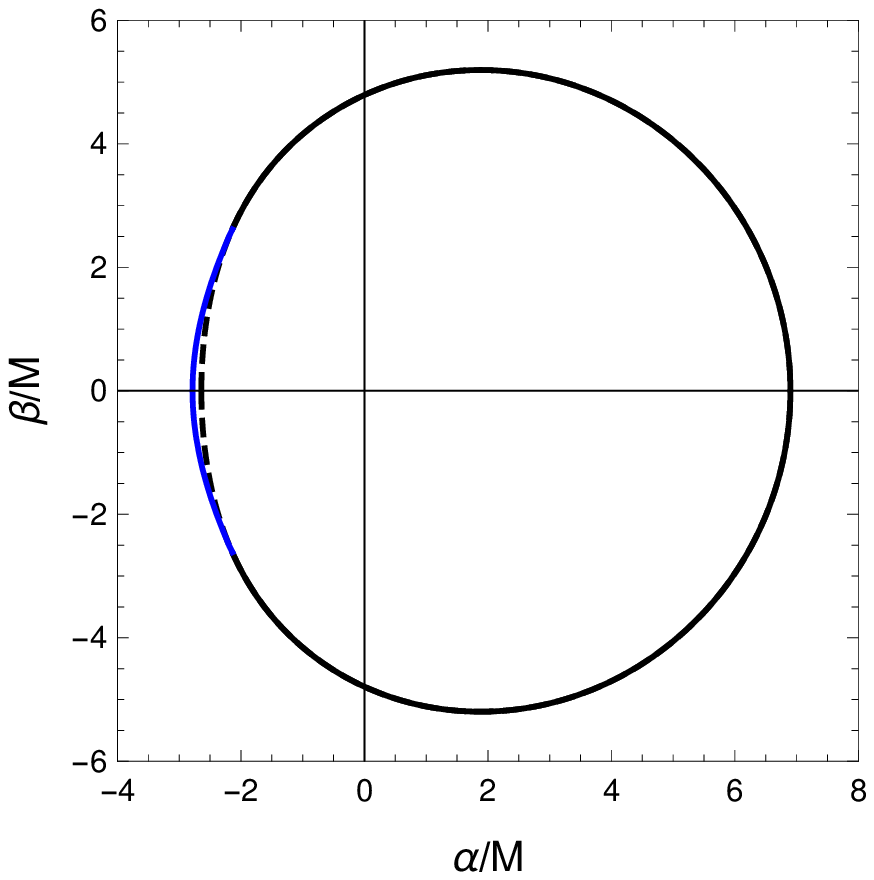}}
\subfigure[$~r_0/r_{+}=1.35$]{\includegraphics[scale=0.68]{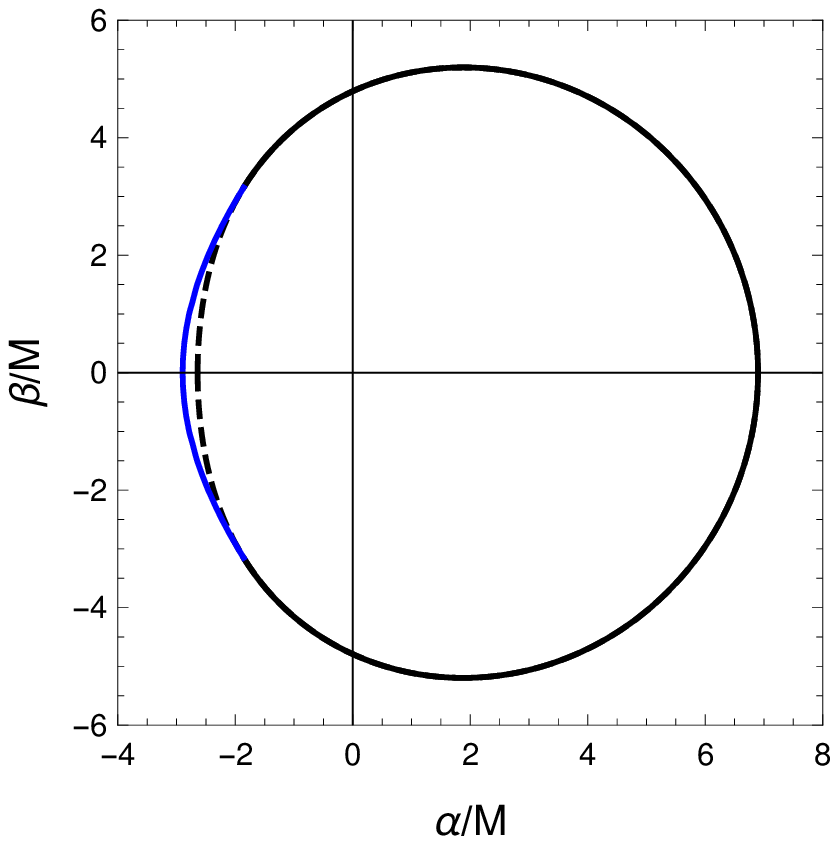}}
\caption{Shadows cast by the black hole alternatives (solid black plus solid blue) and the Kerr black hole (black dashed). The solid blue part of the shadow contour is due to $R(r_0)=0$. Here, the spin is $a_*=a/M=0.94$ and $\theta_0=\pi/2$. For these parameters, the shadow of the Kerr black hole has $r_{ph,min}/r_+=1.063$. Note that the shadow of the object deviate from that of the Kerr black hole only when $r_0>r_{ph,min}$.}
\label{fig:shadowHCO}
\end{figure}

\begin{figure}[h]
\centering
\includegraphics[scale=1.0]{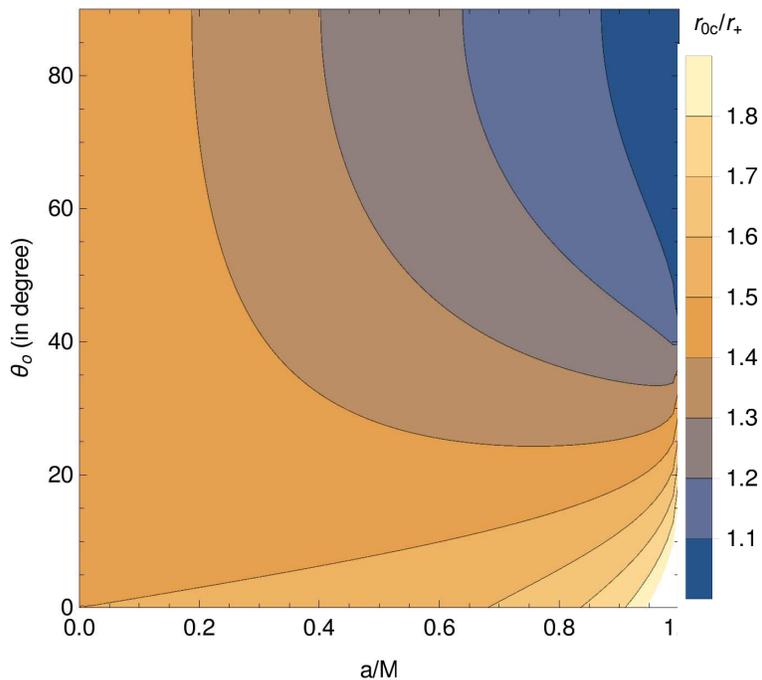}
\caption{Parameter region showing the critical radius $r_{0c}$ of the horizonless alternative above which its shadow starts deviating from that of the Kerr black hole.}
\label{fig:r0c}
\end{figure}

Figure \ref{fig:rph} shows the minimum and the maximum of the unstable photon orbits which take part in the shadow in the Kerr black hole background for a given spin and the observer inclination angle. Figure \ref{fig:shadowHCO} shows some shadows of a horizonless alternative. Note that the shadow of the object deviate from that of the Kerr black hole only when $r_0>r_{ph,min}$. Therefore, we can define a critical radius of the object by $r_{oc}=r_{ph,min}$. If the object has a radius greater than this critical value, i.e., if $r_0\geq r_{0c}$, then its shadow deviates from that of the Kerr black hole. Figure \ref{fig:r0c} shows the dependence of the critical radius on the spin and observation angle. Note that, if the radius of the object lies very close to the would-be event horizon of the Kerr black hole (say $r_0<1.2r_+$ for example), then we need both high spin and high observation angle to probe such modification through the shadow.

\section{Constraining the black hole alternatives from the M87$^*$ observation}
\label{sec:results}
We now use the results from M87$^*$ observation and put possible constraint on the Kerr black hole alternatives. For this purpose, we use the average angular size of the shadow and its deformation from circularity. Note that the shadow is perfectly circular for zero spin and start deforming as we increase the spin. With increasing spin, the center of the shadow in the $\alpha-\beta$ plane also starts shifting from the origin. However, as the shadow has reflection symmetry around the $\alpha$-axis, its geometric center $(\alpha_c,\beta_c)$ is given by $\alpha_c=1/A\int \alpha dA$ and $\beta_c=0$, $dA$ being an area element. We use this geometric centre to find out the average shadow radius and deviation from circularity. To this end, we first define an angle $\phi$ between the $\alpha$-axis and the vector connecting the geometric centre $(\alpha_c,\beta_c)$ with a point $(\alpha,\beta)$ on the boundary of a shadow. Therefore, the average radius $R_{av}$ the shadow is given by \citep{Bambi}
\begin{equation}
R_{av}^2=\frac{1}{2\pi}\int_{0}^{2\pi}l^2(\phi)\; d\phi,
\end{equation}
where $l(\phi)=\sqrt{(\alpha(\phi)-\alpha_c)^2+\beta(\phi)^2}$ and $\phi=tan^{-1}(\beta(\phi)/(\alpha(\phi)-\alpha_c))$. Following \cite{EHT1}, we define the deviation $\Delta C$ from circularity as
\begin{equation}
\Delta C=\frac{1}{R_{av}}\sqrt{\frac{1}{2\pi}\int_{0}^{2\pi}(l(\phi)-R_{av})^2\; d\phi}.
\end{equation}
Note that $\Delta C$ is the fractional RMS distance from the average radius of the shadow.

\begin{figure}[h]
\centering
\subfigure{\includegraphics[scale=0.85]{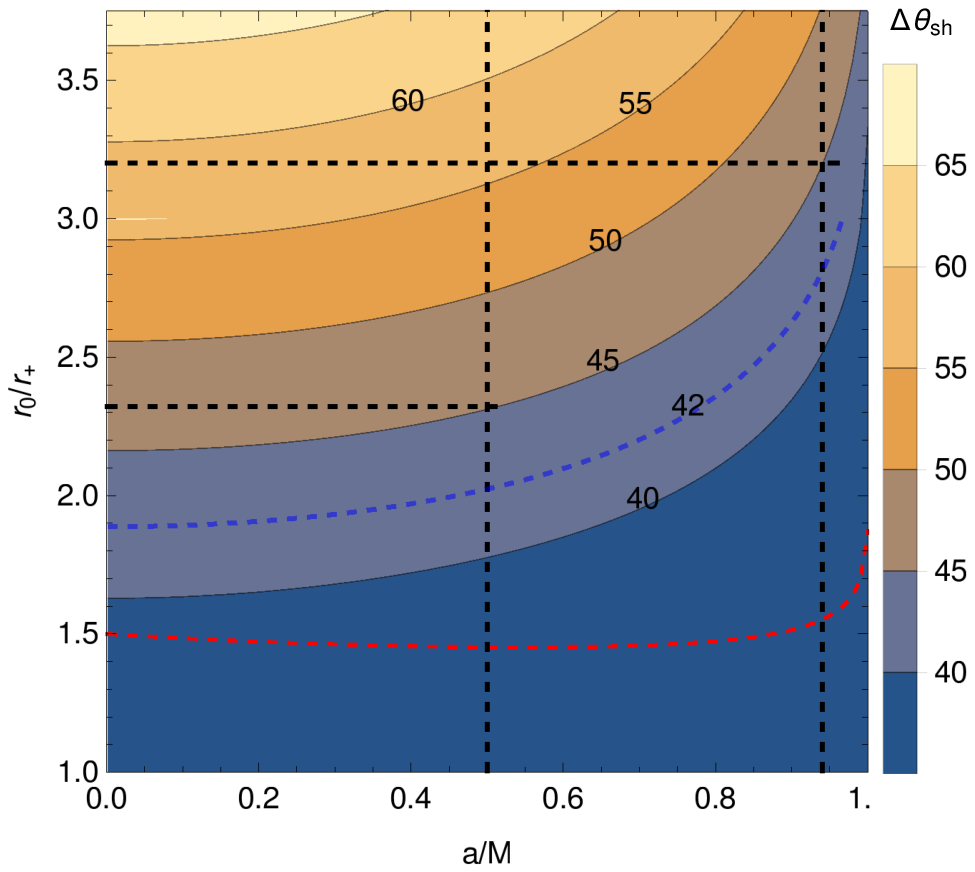}}
\subfigure{\includegraphics[scale=0.85]{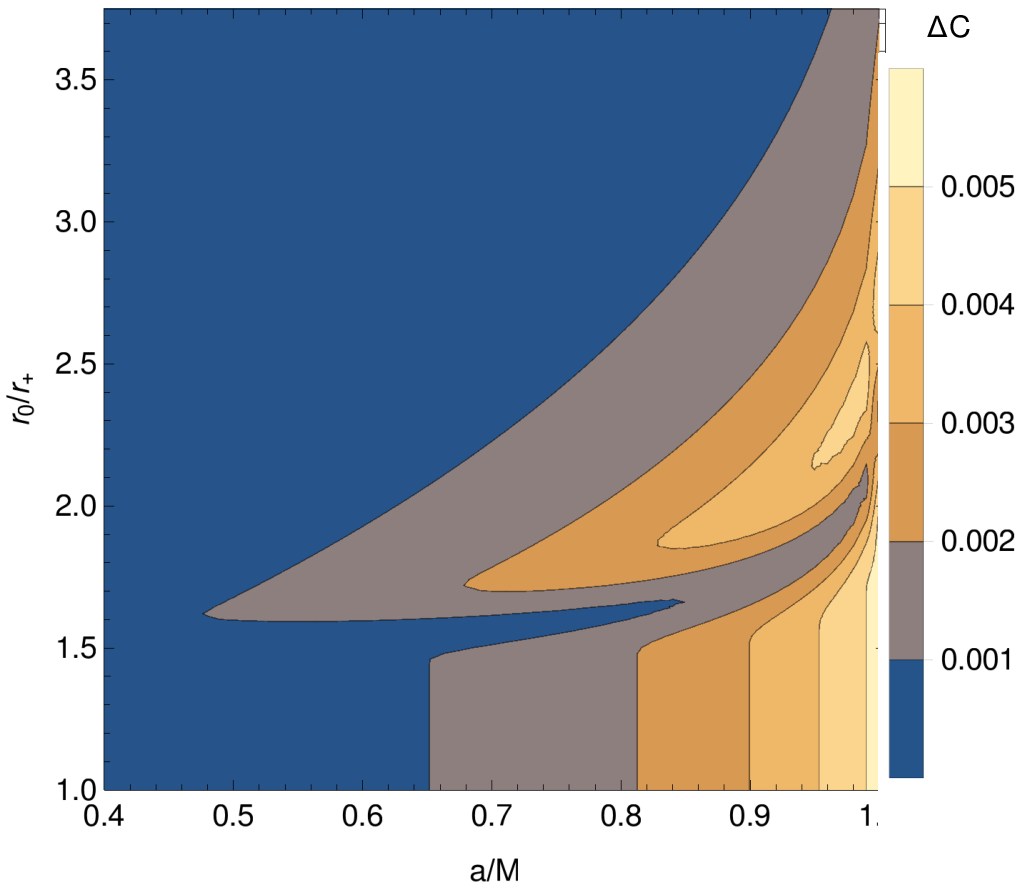}}
\caption{Dependence of the angular size and the deviation of the shadow on $r_0$ and spin. Here, $M=6.5\times 10^9 M_\odot$ and $D=16.8$ Mpc. The region between the two black dashed vertical lines indicate the spin range $0.5\leq a_{*}\leq 0.94$. The two horizontal black dashed lines indicate the range of the maximum value $r_{0,max}$ for this spin range, in order that the angular size is within the maximum value of 45$\mu$as. For a given spin, say for $a/M=0.5$, $r_{0,max}=2.31r_{+}$, indicating that the angular size of the shadow of the black hole alternatives will be consistent with the M87$^*$ observation if $r_0\leq r_{0,max}$. The red dashed line shows the critical value $r_{0c}$ of $r_0$ as discussed in the previous section.}
\label{fig:angularSize}
\end{figure}

The angular diameter of the shadow is given by $\Delta\theta_{sh}=2R_{av}/D$, where $D$ is the distance to M87$^*$. Following \cite{EHT1}, We take $D=(16.8\pm 0.8)$ Mpc and the mass of the object to be $M=(6.5\pm 0.7)\times 10^9 M_\odot$. The inclination angle is taken to be $\theta_o=17^\circ$, which the jet axis makes to the line of sight \citep{EHT1}. With this, we calculate the angular diameter $\Delta\theta_{sh}$ and deviation $\Delta C$ for different $r_0$ and spin $a$. This is shown in Fig. \ref{fig:angularSize}. According to EHT collaboration, the angular size of the observed shadow is $\Delta\theta_{sh}=42\pm 3$ $\mu$as \citep{EHT1} and the spin lies within the range $0.5\leq a_{*}\leq 0.94$. Note form Fig. \ref{fig:angularSize} that, for the spin range  $0.5\leq a_{*}\leq 0.94$, the angular size will be consistent with the M87$^*$ observation, i.e., $\Delta\theta_{sh}$ is between $39$ to $45$ $\mu$as, if the maximum value $r_{0,max}$ of $r_0$ of the black hole alternatives lies within the range $2.31r_{+}\leq r_{0,max}\leq 3.19r_{+}$. If $r_0$ is below this maximum range (i.e. $r_0\leq r_{0,max}$), then the shadow silhouette of the alternatives is consistent with the observed size. For a given spin, say for $a/M=0.5$, $r_{0,max}=2.31r_{+}$, indicating that the angular size of the shadow of the black hole alternatives will be consistent with the M87$^*$ observation if $r_0\leq r_{0,max}$. The deviation of the observed shadow from circularity is reportedly less than $10\%$. From Fig. \ref{fig:angularSize}, note that the deviation of the shadow for the black hole alternatives is consistent with the observed values, for the above mentioned spin range and $r_0<r_{0,max}$.

\begin{figure}[h]
\centering
\includegraphics[scale=0.85]{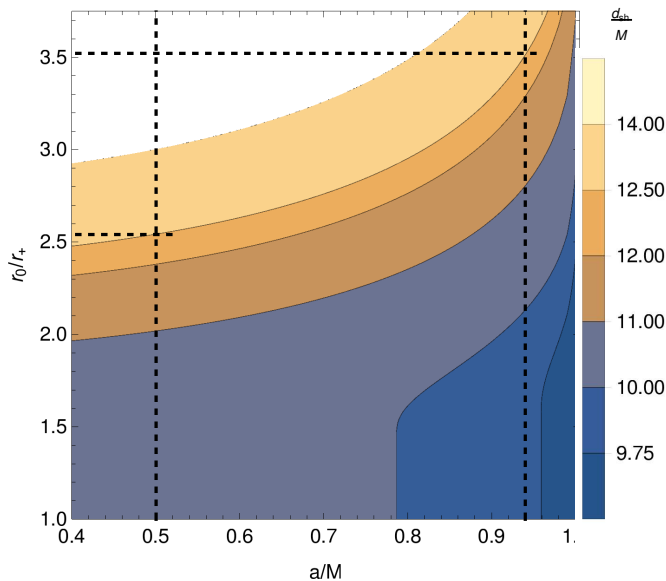}
\caption{Dependence of the size of the shadow on $r_0$ and spin. Here, $M=(6.5\pm 0.7)\times 10^9 M_\odot$ and $D=(16.8\pm 0.8)$ Mpc. The region between the two black dashed vertical lines indicate the spin range $0.5\leq a_{*}\leq 0.94$. The two horizontal black dashed lines indicate the range of the maximum value $r_{0,max}$ for this spin range, in order that the size is within the maximum value of $12.5$. For a given spin, say for $a/M=0.5$, $r_{0,max}=2.54r_{+}$, indicating that the angular size of the shadow of the black hole alternatives will be consistent with the M87$^*$ observation if $r_0\leq r_{0,max}$. The red dashed line shows the critical value $r_{0c}$ of $r_0$ as discussed in the previous section.}
\label{fig:dav}
\end{figure}

Note that, in Fig. \ref{fig:angularSize}, we took $M=6.5\times 10^9 M_\odot$ and $D=16.8$ Mpc and ignored the errors in these quantities. In order to incorporate that, we find the size of the shadow first form the M87$^*$ data. Taking $D=(16.8\pm 0.8)$ Mpc and $M=(6.5\pm 0.7)\times 10^9 M_\odot$, the size of the shadow in dimensionless unit is estimated to be
\begin{equation}
\frac{d_{sh}}{M}=\frac{D\Delta\theta_{sh}}{GM}=11.0\pm 1.5,
\end{equation}
where the errors have been added in quadrature. The above quantity must be equal to $\frac{2R_{av}}{M}$. Figure \ref{fig:dav} shows the the dependence of the shadow size on different parameters. Note that, after incorporating the errors in $M$ and $D$, the range of the maximum value of $r_0$ is modified to $2.54r_{+}\leq r_{0,max}\leq 3.51r_{+}$.

\section{Conclusions}
\label{sec:conclusion}
In this paper, we have considered two viable alternatives to the Kerr black hole. The first one is a horizonless compact object (HCO) having radius $r_0$ and exterior Kerr geometry. The second one is a rotating generalisation of the recently obtained one parameter ($r_0$) static metric by Simpson and Visser. We have studied shadows cast by these alternatives and compared them with the observed shadow of the supermassive compact object M87$^{*}$, thus constraining the parameter $r_0$ of the alternatives. We find that, for the mass, inclination angle and the angular diameter of the shadow of M87$^{*}$ reported by the EHT collaboration, the maximum value of the parameter $r_0$ must be in the range $2.54r_{+}\leq r_{0,max}\leq 3.51r_{+}$ for the dimensionless spin range $0.5\leq a_{*}\leq 0.94$, with $r_{+}$ being the outer horizon radius of the Kerr black hole at the corresponding spin value. We conclude that these black hole alternatives having $r_0$ below this maximum range (i.e. $r_0\leq r_{0,max}$) is consistent with the size and deviation from the circularity of the observed shadow of M87$^{*}$. An increase
in precision in the EHT data should improve the constraint that we have reported here.

\begin{appendices}

\section{Rotating Simpson-Visser metric}
\label{sec:appendix}
The Newman-Janis algorithm is a set of steps to construct a stationary, axisymmetric spacetime beginning from a static and spherically symmetric spacetime \cite{NJ1,NJ2}.  Staring from a spherically symmetric, static spacetime written as
\begin{equation}\label{metr1}
  ds^2=-f(r)dt^2 + \frac{dr^2}{g(r)}+ h(r)(d\theta^2+\sin^2\theta d\phi^2)~,
\end{equation}
we make a coordinate transformation to get the metric in the advanced null coordinate defined as
\begin{equation}
    du=dt-\frac{dr}{\sqrt{fg}} ~~~.
\end{equation}
Then the above metric of eq.(\ref{metr1}) in these coordinates becomes
\begin{equation}
    ds^2=-f(r)du^2-2\sqrt{\frac{f}{g}}dudr + h(r)(d\theta^2+\sin^2\theta d\phi^2) ~.
\end{equation}
The next step is to express the inverse metric $g^{\mu\nu}$ in a null tetrad basis in the form
\begin{equation}
g^{\mu\nu} = -l^{\mu}n^{\nu}-l^{\nu}n^{\mu}+ m^{\mu}\Bar{m}^\nu+m^{\nu}\Bar{m}^\mu,
\end{equation}
where the null tetrad is $Z^{\mu}_{\nu}=\big(l^{\mu},n^{\mu},m^{\mu},\Bar{m}^{\mu}\big)$, and a bar denotes complex conjugation. The tetrad vectors are given by
\begin{equation}
l^\mu=\delta^\mu_r, \hspace{0.2cm} n^\mu=\sqrt{\frac{g}{f}}\delta^\mu_u-\frac{g}{2}\delta^\mu_r, \quad m^\mu=\frac{1}{\sqrt{2h}}\left(\delta^\mu_\theta+\frac{i}{\sin\theta}\delta^\mu_\phi\right).
\end{equation}

Now, we perform a complex transformation given by
\begin{equation}
r \rightarrow r^{\prime} = r+ ia\cos\theta~,~~\text{and}~,~~  u \rightarrow u^{\prime} = u- ia\cos\theta~,
\end{equation}
so that, after the complex transformation, the new tetrad becomes
\begin{equation}
    l'^{\mu}=\delta^{\mu}_{r} ,   
    n'^{\mu}=\sqrt{\frac{G(r,\theta)}{F(r,\theta)}}\delta^{\mu}_{u}-\frac{G(r,\theta)}{2}\delta^{\mu}_{r}~~~,
    ~~~m'^{\mu}=\frac{1}{\sqrt{2H(r,\theta)}}\Big(i a\sin\theta(\delta^{\mu}_{u}-\delta^{\mu}_{r})+\delta^{\mu}_{\theta}+\frac{i}{\sin\theta}\delta^{\mu}_{\phi}\Big)~.
\end{equation}
Here, $F(r,\theta)$, $G(r,\theta)$ and $H(r,\theta)$ are the complexified form of the functions $f(r)$, $g(r)$ and $h(r)$ respectively. At this stage, the metric constructed using the new tetrad obtained after complexification contain non-diagonal components other than $g_{t\phi}$. Therefore, the last step is to write the metric in Boyer-Lindquist form (where the only nonzero off diagonal term is $g_{t\phi}$) using the coordinate transformation
\begin{equation}
    du=dt^{\prime}+\chi_{1}(r)dr, ~~~ d\phi =d\phi^{\prime}+\chi_{2}(r)dr~.
\end{equation}
We must ensure that the two functions $\chi_{1}(r)$ and $\chi_{2}(r)$ are solely functions of $r$ to define the global transformations. 
Here, $\chi_{1}(r), \chi_{2}(r)$ are given by \citep{shaikh_NJ}
\begin{equation}
    \chi_{1}(r)=-\frac{\sqrt{\frac{G(r,\theta)}{F(r,\theta)}}H(r,\theta) + a^2\sin^2\theta}{G(r,\theta)H(r,\theta)+ a^2\sin^2\theta}~,~~\\
 \chi_{2}(r)=-\frac{a}{G(r,\theta)H(r,\theta)+ a^2\sin^2\theta}~.
\end{equation}
The above steps can be used to derive the Kerr metric from the Schwarzschild metric provided we use the following rule for complexifying the metric functions,
\begin{equation}
    \frac{1}{r}\rightarrow \frac{1}{2}\Big(\frac{1}{r^{\prime}}+ \frac{1}{\bar{r}'}\Big) =  \frac{r}{\Sigma} ~~~~\text{and}~~~ r^2 \rightarrow r^{\prime}\bar{r}'= \Sigma~,
\end{equation}
where $\Sigma= r^2+a^2\cos^2\theta$. Now to apply this algorithm to the SV metric, we write down the metric functions as
\begin{equation}
f(r)=g(r)=1-\frac{2m(r)}{r^2+r_0^2},\;\; h(r)=r^2+r_0^2~,
\end{equation}
where $m(r)=M\sqrt{r^2+r_0^2}$ and use $r^2 \rightarrow r^{\prime}\bar{r}'= \Sigma$ to obtain the complexified functions as
\begin{equation}
    F(r,\theta)=G(r,\theta)= 1-\frac{2 M \sqrt{r^2+r_{0}^2}}{\Sigma+r_{0}^2},\;\; H(r,\theta) = \Sigma + r_{0}^2~.
\end{equation}
With this $\chi_{1}(r)$ and $\chi_{2}(r)$ become
\begin{equation}
\chi_1(r)=-\frac{r^2+r_0^2+a^2}{r^2+r_0^2+a^2-2m(r)}~,~~ \chi_2(r)=-\frac{a}{r^2+r_0^2+a^2-2m(r)}~.
\end{equation}
Finally, the metric turns out to be
\begin{eqnarray} 
ds^2 &=&-\left(1-\frac{2M\sqrt{r^2+r_0^2}}{\Sigma}\right)dt^2-\frac{4Ma\sqrt{r^2+r_0^2} \sin^2\theta}{\Sigma}dt d\phi +\frac{\Sigma}{\Delta}dr^2 \nonumber \\
& & +\Sigma d\theta^2 + \left( r^2+r_0^2+a^2+\frac{2Ma^2\sqrt{r^2+r_0^2} \sin^2\theta}
{\Sigma}\right) \sin^2\theta d\phi^2~,
\end{eqnarray}
\begin{equation}
 \Sigma=r^2+r_0^2+a^2 \cos^2\theta, \quad   \Delta=r^2+r_0^2-2Mr+a^2~.
\end{equation}
\end{appendices}

\end{document}